# Study of Time Evolution for Approximation of Two-Body Spinless-Salpeter Equation In Presence of Time-Dependent Interaction

*Hadi Sobhani[1*] and Hassan Hassanabadi*

[1]*Physics Department, Shahrood University of Technology,*

*Shahrood, Iran*

*P. O. Box: 3619995161-316*

*\* Email:hadisobhani8637@gmail.com*

**Abstract**

We approximate the two-body spinless Salpeter equation with the one which is valid in heavy quarks limit. We consider the resulting semi-relativistic equation in a time-dependent formulation. We use the Lewis- Riesenfeld dynamical invariant method and series solution to obtain the solutions of the differential equation. We have also done some calculations in order to derive the time evolution operator for the considered problem.

1. Introduction

Salpeter equation describes the bound states of relativistic systems in a covariant formalism [1]. Until now, the equation has been solved by different approaches. Wick transformed the relative momentum into an Euclidean vector to avoid the propagator singularity and thereby obtained some considerable mathematical theorems to solve the equation [2]. A very economical approach to deal with the equation is the Deser- Gilbert-Sudarshan-Ida representation as used by Wick and Cutkosky [3]. Other useful techniques are well addressed in Refs. [4-6]. Salpeter equation, on the other hand, can be considered as the generalization of the nonrelativistic Schrödinger equation into the relativistic regime [7]. A very unappealing characteristics of the equation is its nonlocal nature [8]. Jacquemin et al. calculated the Salpeter the vertical excitation energies for the set of 28 molecules constituting the well-known Thiel's set [9]. Mainland considered the equation in the ladder approximation when the bound-state energy is zero [10]. Eichmann *et al.* presented a numerical solution of the four-quark Salpeter equation for a scalar tetraquark [11]. Shao et al. established the equivalence between Salpeter eigenvalue problems and real Hamiltonian eigenvalue problems [12] Mishima et al. investigated Salpeter equation by employing the Dyson-Schwinger method together with the Munczek-Nemirovsky model [13]. Owen and Barrett used quantum electrodynamics and the Salpeter equation to calculate the bound state energies for a two-particle system comprised of a spin-0 and spin-1/2 particle [14]. Carbonell and Karmanov calculated the transition form factor for electro-disintegration of a two-body bound system in the Salpeter framework

[15]. It should be emphasized that the spinless saltpeter equation (SSE), originates from the Bethe-Salpeter equation by making some simplifications to the equation and neglecting spin degrees of freedom [16-19].

On the other hand, one of the most important problems in quantum mechanics is the time-evolution of the quantum systems. When the time evolution enters the problem, we have to deal with a partial differential equation which is definitely more complicated than the ordinary counterparts. In most cases, the exact analytical techniques fail and we have to use approximate methods for our real physical problems. Till now, a variety of techniques have been applied to the field including the path integral [20], dynamical invariant [21-23] and Gaussian wave packet [24]. Albeverio and Mazzucchi considered Schrödinger equation with a time-dependent quadratic plus quartic Hamiltonian and using Feynman path integral representation [25]. Ibarra-Sierra et al. used the Lie-algebraic technique and solved the time-dependent harmonic oscillator and the bi-dimensional charged particle in time-dependent electromagnetic fields [26]. In the Ref. [27], linear invariants and the dynamical invariant method are used to obtain the exact solutions of the Schrödinger equation for the generalized time-dependent forced harmonic oscillator. Choi used the dynamical invariant method to solve the time-dependent Hamiltonian including quadratic, inverse quadratic, and $(\frac{1}{x})p + p(\frac{1}{x})$ terms [28]. Interesting aspects of Gaussian wave packet technique, which is useful in the field, can be found in Refs. [29-31].

Here, we are going to consider a time-dependent approximation of the SSE via dynamical invariant method originally proposed by Lewis and Riesenfeld [22]. In Sec. 2, we introduce our time-dependent semi-relativistic equation. Sec. 3 reports the solutions of the problem. The calculation of the evolution operator for SSE are appeared in Sec. 4. And the last part shows the concluding.

2. **Time-Dependent Hamiltonian of Semi-Relativistic Spinless-Salpeter Equation**

The Hamiltonian of two-body SSE, in the center of mass framework, has the form

$$H(x,t) = \sum_{i=1}^{2}\sqrt{p^2 + m_i^2} + V(x,t) - \sum_{i=1}^{2} m_i , \ (c=1) \quad (1)$$

Using the binomial expansion, the inverse square term can be written as [32]

$$\sum_{i=1}^{2}\left(p^2 + m_i^2\right)^{\frac{1}{2}} \approx m_1 + m_2 + \frac{p^2}{2\mu} + \frac{p^4}{8\eta^3} + ..., \quad (2)$$

with $\mu = \frac{m_1 m_2}{m_1 + m_2}$ and $\eta = \mu\left(\frac{m_1 m_2}{m_1 m_2 - 3\mu^2}\right)^{\frac{1}{3}}$. Before proceeding further, it should be emphasized that the above approximation is only valid for heavy quark systems. Substituting Eq. (2) into (1) and considering the relative interaction $V(x,t) = f(t)x$, brings the Hamiltonian into the form

$$H(t) = \frac{p^4}{8\eta^3} + \frac{p^2}{2\mu} + f(t)x, \quad (3)$$

The existence of $f(t)$ term in the Hamiltonian prevents Eq. (3) to appear in the form of a known eigenvalue problem and we therefore introduce the Lewis- Riesenfeld dynamical invariant method in the forthcoming section.

### 3. Lewis– Riesenfeld Dynamical Invariant Method and Time-Evolution of the Problem

In 1969, in an article published by Lewis- Riesenfeld, appeared a theory which helps us to treat explicitly the time-dependent systems without using directly the time-dependent Schrodinger equation. According to Lewis- Riesenfeld method [18], there is an invariant Hermitian operator described by

$$\frac{dI(t)}{dt} = \frac{\partial I(t)}{\partial t} + \frac{1}{i\hbar}[I(t), H(t)] = 0, \quad (4)$$

If Eq. (4) acts on an arbitrary Ket from the left, it results in

$$\{\frac{1}{i\hbar}(IH - HI) + \frac{\partial I}{\partial t}\}|\Psi\rangle = 0,$$

$$\frac{\partial I}{\partial t}|\Psi\rangle + I\frac{\partial|\Psi\rangle}{\partial t} = \frac{1}{i\hbar}HI|\Psi\rangle,$$

$$i\hbar\frac{\partial(I|\Psi\rangle)}{\partial t} = H(t)(I|\Psi\rangle). \quad (5)$$

This implies that action of the invariant operator on the Ket, satisfies the time-evolution equation, too.

So if we can find the dynamical invariant in such a way, we can obtain the wave function with the aim of the dynamical invariant eigen functions.

In order to find the explicit form of the invariant, we suggest

$$I(t) = A(t)p^4 + B(t)p^3 + C(t)p^2 + D(t)p + E(t)x + F(t), \quad (6)$$

where the capital letters are arbitrary functions of time. Although Eq. (6) has a crude form, but by substituting Eqs. (6) and (3) into Eq. (4) and some algebraic process, we can get to

$$\dot{A}(t)p^4 + \{\dot{B}(t) - A(t)f(t) + \frac{E(t)}{2\eta^3}\}p^3 + \{\dot{C}(t) - 3B(t)f(t)\}p^2 + \{\dot{D}(t) - 2C(t)f(t)\}p$$
$$+ \dot{E}(t)x + \dot{F}(t) - D(t)f(t) = 0. \quad (7)$$

Eq. (7) is relation which gives us the constraints on the time-dependent coefficients in Eq. (6). Using Eq. (7) the below relations are derived

$$\dot{A}(t) = 0, \quad \dot{E}(t) = 0, \quad (8)$$

$$\dot{F}(t) - D(t)f(t) = 0, \quad (9)$$

$$\dot{B}(t) - A(t)f(t) + \frac{E(t)}{2\eta^3} = 0, \quad (10)$$

$$\dot{C}(t) - 3B(t)f(t) = 0, \quad (11)$$

$$\dot{D}(t) - 2C(t)f(t) = 0, \quad (12)$$

Eqs. (8) to (12) immediately yield

$$A = \text{constant}, \quad E = \text{constant}, \quad (13)$$

$$B(t) = \int [Af(t) + \frac{E}{2\eta^3}]dt + \text{constant}, \quad (14)$$

$$C(t) = \int 2B(t)f(t)dt + \text{constant}, \quad (15)$$

$$D(t) = \int [2C(t)f(t) + \frac{E}{m}]dt + \text{constant}, \quad (16)$$

$$F(t) = \int D(t)f(t)dt + \text{constant}. \quad (17)$$

Therefore, given an explicit form of $f(t)$, the explicit form of the invariant can be determined. Let us now return to the eigenfunction. Because there is no operator term with respect to time in the dynamical invariant operator form, we can consider these terms as constants with respect to the coordinates. The eigenvalue problem for the time-dependent operator is

$$I\Phi = \lambda\Phi, \quad (18)$$

Where $\Phi$ and $\lambda$ are the time-dependent eigenfunction and time-independent eigenvalue, respectively. Inserting the explicit form of the invariant and utilizing $p = -i\hbar\frac{d}{dx}$, we get

$$\hbar^4 A \frac{d^4\Phi}{dx^4} - i\hbar^3 B(t)\frac{d^3\Phi}{dx^3} - C(t)\hbar^2 \frac{d^2\Phi}{dx^2} - i\hbar D(t)\frac{d\Phi}{dx} + Ex\Phi + F(t)\Phi = \lambda\Phi, \quad (19)$$

To find the solution of Eq. (19), we propose a series solution of the form $\Phi = \sum_{n=0}^{\infty} a_n x^n$. Substitution of the latter in Eq. (19) results in

$$\hbar^4 A \sum_{n=0}^{\infty} n(n-1)(n-2)(n-3)a_n x^{n-4} - i\hbar^3 B(t)\sum_{n=0}^{\infty} n(n-1)(n-2)a_n x^{n-3} - C(t)\hbar^2 \sum_{n=0}^{\infty} n(n-1)a_n x^{n-2}$$

$$-i\hbar D(t)\sum_{n=0}^{\infty} na_n x^{n-1} + E\sum_{n=0}^{\infty} a_n x^{n+1} + F(t)\sum_{n=0}^{\infty} a_n x^n = \lambda \sum_{n=0}^{\infty} a_n x^n, \quad (20)$$

which gives the recurrence relation as

$$\hbar^4 A \frac{(n+5)!}{(n+1)!} a_{n+5} - i\hbar^3 B(t) \frac{(n+4)!}{(n+1)!} a_{n+4} - C(t)\hbar^2 \frac{(n+3)!}{(n+1)!} a_{n+3} - i\hbar D(t)(n+2)a_{n+2} + E a_n + (F(t)-\lambda)a_{n+1} = 0, \quad (21)$$

Finally, to write the wave function we assume $\Psi = \kappa(t)\Phi$. Using the time evolution equation of the wave function, we may have

$$i\hbar \frac{\partial \Psi}{\partial t} = H\Psi, \rightarrow i\hbar \frac{\partial(\kappa(t)\Phi(x,t))}{\partial t} = H(\kappa(t)\Phi(x,t)),$$

$$\rightarrow i\hbar(\frac{\partial \kappa(t)}{\partial t}\Phi(x,t) + \kappa(t)\frac{\partial \Phi(x,t)}{\partial t}) = \kappa(t)H(\Phi(x,t)),$$

$$i\hbar(\frac{1}{\kappa(t)}\frac{\partial \kappa(t)}{\partial t} + \frac{\partial \Phi(x,t)}{\partial t}) = H\Phi(x,t),$$

$$\rightarrow i\hbar \frac{1}{\kappa(t)}\frac{\partial \kappa(t)}{\partial t} = H\Phi(x,t) - i\hbar \frac{\partial \Phi(x,t)}{\partial t}.$$

$$\ln \kappa(t) = \int \left( \frac{H\Phi(x,t)}{i\hbar} - \frac{\partial \Phi(x,t)}{\partial t} \right) dt,$$

$$\rightarrow \kappa(t) = \exp[\int (\frac{H\Phi}{i\hbar} - \frac{\partial \Phi(x,t)}{\partial t}) dt], \quad (22)$$

we set the integration constant equal to zero. So Using $\kappa(t)$ and $\Phi(x,t)$, the wave function can be written as $\Psi(x,t) = \kappa(t)\Phi(x,t)$.

In the next section, another aspect of the time evolution of the considered system will be investigated.

### 4. Time Evolution Operator for Time-Dependent SSE

Other aspect of the time evolution study of a system is having the time evolution operator. Time Evolution operator, is an unitary operator that

$$\Psi(x,t) = U(t)\Psi(x,0), \quad (23)$$

where $U(t)$ should satisfy

$$i\hbar \frac{dU(t)}{dt} = HU(t). \quad (24)$$

Since the time evolution operator is unitary, using this property and multiplying Eq. (24) by $U^{-1}(t)$ from the right, it changes

$$i\hbar \frac{dU(t)}{dt}U^{-1}(t) = H. \quad (25)$$

Eq. (25) helps us to find appropriate evolution operator for Eq. (3). In order to obtain the operator we should assume an ansatz [33]

$$U(t) = \exp\left[\gamma_1(t)p^4 + \gamma_2(t)p^3 + \gamma_3(t)p^2 + \gamma_4(t)p + \gamma_5(t)x + \gamma_6(t)\right], \quad (26)$$

in which $\gamma_i(t), (i = 1,2,3,4,5,6)$ should be determined. In order to do this we should insert Eq. (26) into Eq. (25), which yields

$$i\hbar \frac{dU(t)}{dt} U^{-1}(t) = i\hbar[\dot{\gamma}_1(t)p^4 + \dot{\gamma}_2(t)p^3 + \dot{\gamma}_3(t)p^2 + \dot{\gamma}_4(t)p$$
$$+ \dot{\gamma}_5(t)e^{\gamma_1(t)p^4 + \gamma_2(t)p^3 + \gamma_3(t)p^2 + \gamma_4(t)p} \, x \, e^{-\gamma_4(t)p - \gamma_3(t)p^2 + \gamma_2(t)p^3 + \gamma_1(t)p^4} + \dot{\gamma}_6(t)],$$

Using Baker–Hausdorff formula, Eq. (27) can be rewritten simpler as

$$i\hbar \frac{dU(t)}{dt} U^{-1}(t) = i\hbar[\dot{\gamma}_1(t)p^4 + p^3\left(\dot{\gamma}_2(t) - 4i\hbar\gamma_1(t)\dot{\gamma}_5(t)\right) + p^2\left(\dot{\gamma}_3(t) - 3i\hbar\dot{\gamma}_5(t)\gamma_2(t)\right) +$$
$$p\left(\dot{\gamma}_4(t) - 2i\hbar\dot{\gamma}_5(t)\gamma_3(t)\right) + x\dot{\gamma}_5(t) - i\hbar\dot{\gamma}_4(t) + \dot{\gamma}_6(t)], \quad (28)$$

and comparing Eq. (28) and (3), we find out

$$\gamma_1(t) = \frac{-it}{8\hbar\eta^3} + \text{constant}, \quad (29)$$

$$\gamma_2(t) = 4\int f(t)\gamma_1(t)dt + \text{constant}, \quad (30)$$

$$\gamma_3(t) = \frac{-3i}{\hbar}\int f(t)\gamma_2(t)dt - \frac{it}{2\hbar\mu} + \text{constant}, \quad (31)$$

$$\gamma_4(t) = 2\int f(t)\gamma_3(t)dt + \text{constant}, \quad (32)$$

$$\gamma_5(t) = \frac{-i}{\hbar}\int f(t)dt + \text{constant}, \quad (33)$$

$$\gamma_6(t) = i\hbar\int \gamma_4(t)dt + \text{constant}, \quad (34)$$

By these parameters the evolution will be given. This section shows that having the explicit form of $f(t)$ and $f(0)$ we can find the wave function at the $t = 0$. Also, using the wave function and the time evolution operator which was derived during this section we can obtain the wave function at any time.

**Conclusion**

We studied an approximation of the semi-relativistic spinless Saltpeter equation. We used Lewis-Riesenfeld dynamical invariant to investigate time evolution of the Hamiltonian. The explicit form of the dynamical invariant was derived. In order to obtain eigenfunction of this invariant we dealt with a differential equation with variable coefficients. We had to suggest the solution in series form. The wave function was obtained by using the eigenfunction of the dynamical invariant and appropriate evolution operator. This kind of investigations opens new insight for possible further studies to study the dynamical properties of heavy mesons.

**Acknowledgment**

It is a great pleasure for the authors to thank the referee because of the helpful comments.

**References**

1. H. Bethe, E. Salpeter ,*Physical Review* **84** (6): 1232, (1951).
2. G. C. Wick, Phys. Rev. 96, 1124, (1954).
3. R.E.Cutkosky, Phys.Rev.96,1135(1954).
4. G. Wanders and Helv, Phys .Acta 30, 417 (1957).
5. M. Ida and K. Maki, Progr. Theoret. Phys. (Kyoto) 26, 470 (1961).
6. I .Sato, J.Math.Phys.4,24(1963).
7. S. Hassanabadi, and et al., Chin. Phys. B. 22, 060303 (2013).
8. W.Lucha and F.F. Schoberl, Phys. Rev. C. 56,3369 (1994).
9. D. Jacquemin and et al. Chem. Theory Comput.11 (7), 3290–3304 (2015).
10. G. B. Mainland, Few-Body Syst 56, 197–218, (2015).
11. G. Eichmann and et al. arXiv:1507.05022 (2015).
12. M. Shao and et al., arXiv:1501.03830v3 (2015).
13. G.Mishima and et al. Phys. Rev. D 91, 076011 (2015).
14. D. Owen and R. Barrett, arXiv:1505.06809v1(2015).
15. J. Carbonell and V.A. Karmanov, Phys. Rev. D 91, 076010 (2015).
16. H. Hassanabadi and et al., Chin. J. Phys., 50,5 (2012)
17. S. Hassanabadi and et al.,Chin. Phys. B, 22, 6, 060303 (2013).
18. S. Hassanabadi and et al., Chin. Phys. C, 37, 8, 083102 (2013).
19. S. Zarrinkamar and et al., Few-Body Syst 54:2001–2007 (2013).
20. H. G. Oh and et al., Phys. Rev. A 39, 5515 (1989)
21. M. Maamache and H. Choutri, J. Phys. A 33, L6203 (2000)
22. H.R. Jr.Lewis and W.B.Rienesfeld, J. Math. Phys.10, 1458 (1969)

23. I.Guedes, Phys. Rev. A68, 016102 (2003)

24. E. J. Heller, Chem. Phys. Lett. 34, 321 (1975)

25. S. Albeverio and S. Mazzucchi, Jou.Func. Ana. 238, 471–488, (2006).

26. V. G. Ibarra-Sierra and et al., Ann. Phys. 00, 1–32 (2015).

27. A. L. de Lima and et al., Ann. Phys. 323, 2253–2264 (2008).

28. J. R. Choi, Int. J. Theo. Phys. 42, 4 (2003).

29. I. A. Pedrosa and et al. Phys. DOI: 10.1139/cjp-2014-0553 (2015).

30. S. Menouar and M. Maamacheand J. R. Choi, Chi.Jou. Phys. 49, 4, 871-876 (2011).

31. S. Menouar and et al., J. Kor. Phys. Soc., 58, 1, 154-157 (2011).

32. S. Hassanabadi and et al, Chin. Phys. C ,37, 8, 083102 (2013).

33. J. Wei and E. Norman, J. Math. Phys. 4, 575 (1963).